\documentclass[12pt,preprint]{aastex}
\usepackage{amsmath}
\newcommand{\reducedI} {\overset{-} {\smash {\mathbf{I}}}}
\DeclareMathOperator{\trace}{tr}%
\DeclareMathOperator{\diagonal}{diag}%
\shorttitle{Polarizable Vacuum Cosmology}%
\shortauthors{Ibison}%
\begin{document}
\title{An Investigation of the Polarizable Vacuum Cosmology}%
\author{M. Ibison}%
\affil{Institute for Advanced Studies at Austin}%
\affil{4030 West Braker Lane, Suite 300, Austin, Texas 78759}%
\email{ibison@earthtech.org}%
\email{submitted to ApJ, January 2004}
%
\begin{abstract}
%
The basic cosmological predictions of a `polarizable-vacuum' theory of gravity due to Dicke, predating the scalar-tensor theory, are
investigated and tested for their compliance with current observations. A Friedmann-like equation is derived for the theory that
differs in a significant way from that of GR, for which it is demonstrated that a big bang is not a solution: the initial scale factor
may be small but not zero, and must have zero first derivative. Detailed graphs of the predictions of the theory for combinations of
the density parameters are given. It is shown that these predictions are compatible with current estimates for the age and
deceleration parameter, and the maximum observable redshift, provided the matter density $\Omega_m$ is close to the luminous matter
density, i.e. provided there is not a significant amount of missing mass. For these best fit parameters, the theory predicts a maximum
initial temperature that is too small for BBN and probably too small to support an era of radiation domination. The theory predicts a
radiation power from a binary system that is 2/3 that predicted by GR, and so incompatible with observed orbital decay rate of PSR
1913 + 16. There is some discussion of possible developments, including reconciliation through accommodation of vacuum anisotropy, and
of the relation between PV and the Yilmaz theory and the quasi-steady-state cosmology.
\end{abstract}
%
\section{Introduction}
%
That the action of gravity on light can be interpreted as a variable vacuum refractive index has been a persistent theme in the
literature - \citet{Eddington,Gordon,Balazs,Skrotskii,Plebanski,Atkinson1,Atkinson2,Bertotti,Winterberg,Volkov,deFelice,Evans1,Evans2}
- and in some of the standard texts on GR - \citet{L&L,Weyl,Moller,Fock}. The condition that gravitational action on light be
\emph{exactly} expressible in terms of a variable scalar refractive index $n(\mathbf{x},t)$ is just that the metric can be put into
isotropic form, whereupon $c/n(\mathbf{x},t)$ is the coordinate speed of light (the proper speed of light is always $c$). Since this
condition is met outside the event horizon of the Schwarzschild solution, and by all the FLRW metrics, the refractive index
interpretation and corresponding mathematics of EM propagation in gravitational fields has wide application. In the event that the
coordinate system cannot be put into isotropic form, \citet{deFelice,Volkov,L&L} have given the general relationship between the
metric tensor and the generalized constitutive relations to be ascribed to the vacuum that will correctly predict the behaviour of the
EM fields. These relations include the requirement, in order that EM radiation suffer no bi-refringence, that the electric
permittivity and magnetic susceptibility are mutually proportional. In summary, these works are based upon the premise that, ignoring
the action upon matter but otherwise without loss of physical accuracy, the gravitational action on EM fields in a given metric can be
interpreted as due to gravitationally-modulated \emph{vacuum polarizability}.

\citet{Wilson} was probably the first to suggest extension of the (specifically) refractive index paradigm to include matter, by which
means he demonstrated agreement with theory to the level of the Newtonian approximation. Subsequently, \citet{Atkinson1,Atkinson2}
gave the dependence required of the rest mass on the local vacuum refractive index in order that matter precisely follow the geodesics
of the Schwarzschild solution. Interpreted literally, though the details are omitted, the efforts of Wilson and Atkinson amount to the
presumption of an electromagnetic basis for mechanical mass. Supposing this claim can be substantiated, such an approach, however,
does not constitute an electromagnetic theory of gravity for as long as the Einstein action has no electromagnetic explanation.
\citet{Barcelo} have given suggestions for how the latter might emerge from an \emph{effective} action wherein the observable degrees
of freedom are really linearized fluctuations of an invisible \emph{background} field.

In addition to the authors cited above, \citet{Dicke1,Dicke2}, also, was an advocate of the refractive index interpretation of the
action of gravity on light. Like Wilson and Atkinson, Dicke went on to speculate on an electromagnetic basis for mechanical mass,
whereby the polarizable vacuum paradigm could be extended to encompass not only EM but also matter. His work has recently been
re-visited by \citet{Puthoff}. Unlike all previously cited authors apart from Puthoff, however, it is evident in those papers that
Dicke's motivation was to demonstrate the viability of a theory that is electromagnetically motivated from the outset, rather than
simply to forge agreement between GR and a theory of a polarizable vacuum (hereafter, `PV'). An important consequence of that
difference is that Dicke adduces a metric that differs from that of Schwarzschild in the higher order terms of the PPN expansion.
Specifically, if the components of the line element for the latter are written in isotropic coordinates as
\begin{equation} \label{GRSchwarzLineElement}
ds^2=\left(\frac{1-\psi}{1+\psi}\right)^2 c^2dt^2 - \left({1+\psi}\right)^4 d\mathbf{x}^2
\end{equation}
where $\psi = Gm/2rc^2$, then the corresponding PV line element, \citep{Puthoff}, is
\begin{equation} \label{PVSchwarzLineElement}
ds^2=e^{-4\psi} c^2dt^2 - e^{4\psi} d\mathbf{x}^2 \,.
\end{equation}
(Here and throughout, rather than set to 1 through redefinition of the time coordinate, the speed of light $c$ is retained in respect
of the interpretation of $K$ as modifying the vacuum speed in a Euclidean background.) Consistent with this line element, the PV
metric always has the form
\begin{equation} \label{PVLineElement}
ds^2=c^2dt^2/K - K d\mathbf{x}^2
\end{equation}
where $K \equiv K\left(\mathbf{x},t\right)$ is the vacuum polarizability. The specification of PV is complete by assigning to the
$K$-field, on the basis of maximum mathematical simplicity consistent with local Lorentz Invariance, the action of a scalar wave
equation with speed $c/K$.

It is important that though Dicke's reasoning lead him to equations \eqref{PVSchwarzLineElement}) and (\ref{PVLineElement}), it did
not do so unambiguously; his choice rested ultimately on the criterion of mathematical simplicity, given that gravitational action is
fundamentally electromagnetic and mediated by a polarizable vacuum. This is best demonstrated by contrasting the very simple
expressions obtained by PV for the refractive index and mass-dependency on the gravitational field strength parameter $\psi$, for
example, with the results of \citet{Atkinson1}. In the latter, it is recalled, GR effects on EM and matter are driven by the
requirement that the (GR) Schwarzschild metric be reproduced exactly. Atkinson's work achieves the stated aim, but at the price of
cumbersome expressions relating the refractive index and rest-mass to $\psi$, making the correspondence look like a force-fit
\citep{Atkinson1}. In particular, eliminating $\psi$, it can be shown that in the case of the Schwarzschild metric the rest-mass is a
sixth-order polynomial in the refractive index, and cannot, therefore, be given in closed form. \citet{deFelice}, whose aim was to
establish the formal correspondence for the EM fields alone, concludes likewise. Specifically he writes that ``the medium equivalent
approach does not look any simpler than general relativity; on the contrary it may be mathematically more cumbersome and less
elegant''. By contrast, in PV the refractive index and mass-dependencies are both exponentials of $\psi$, and therefore are related by
a simple power.

As noted by \citet{Dicke2}, the Yilmaz theory of gravity \citep{Yilmaz2,Cooperstock,Alley} has in common with PV a line element of the
form (\ref{PVLineElement}) in the particular case that $T_{00}$ is the only non-zero element of the stress energy tensor, i.e. where
the source matter is static, and the pressure / stress and momentum terms are zero \citep{Alley}. Given that Yilmaz originally
conceived of his theory in terms of a polarizable vacuum, \citet{Yilmaz1}, this shared characteristic lends some support to the
`electromagnetic interpretation' of the PV metric. Further connections between the two theories are discussed in section
\ref{sec:developments}.

Though based upon a single scalar function, this remarkably simply theory passes the four standard tests of GR. Yet, to the extent
that PV is driven fundamentally by an EM paradigm, it is a representative of a genre that is quite different from GR. However, apart
from the recent report by Puthoff, it appears not to have been otherwise pursued since Dicke's original publication. The aim of this
document is to fill the gap and thereby permit a more informed assessment of the viability of PV and its domain of applicability. In
the following then, the theory is applied to a few cosmological and astrophysical problems, and its predictions compared with the
data.
%
\section{PV Friedmann equation}
\subsection{Action}
%
Dicke and Puthoff give the classical action for a single massive charge plus EM fields in a polarizable vacuum (in SI units)
\begin{equation} \label{2}
\begin{split}
&L =  - \int {d^4 x} \Biggl\{
        \frac{c^4 } {32\pi GK^2 }
            \left[
                {\left( \nabla K \right)^2  -
                \left( \frac{K} {c}\frac{\partial K} {\partial t} \right)^2 }
            \right] \\
   &+ \left[
            \frac{m_0 c^2 } {\sqrt{K} }\sqrt {1 - \left( \frac{Kv} {c} \right)^2 } + e\left( \phi - \mathbf{v.A} \right) \right]
        \delta ^3\left( \mathbf{x} - \mathbf{x}\left( t \right) \right) \\
   &+ \frac{1} {2} \left[\frac{\mathbf{B}^2 } {K\mu _0 } - K\varepsilon _0 \mathbf{E}^2  \right] \Biggr\}
\end{split}
\end{equation}
where $ {\mathbf{x}}\left( t \right)$ is the particle trajectory, ${\mathbf{v}} = d{\mathbf{x}}\left( t \right)/dt$ the particle
velocity, and $v$ its speed. If in the integrations one takes $dx_0 = cdt$, then the action (equation (\ref{2}) and subsequently) is
$c$ times its usual definition. The particle mass $m_0$ is independent of $K$. The EM field strengths $\mathbf{E}$ and $\mathbf{B}$
have the usual relation to the potentials, $ \mathbf{E} =  - \left( \nabla \phi +
\partial {\mathbf{A}}/ \partial t \right) ,\quad {\mathbf{B}} = \nabla  \times {\mathbf{A}}.
$ The contribution from the $K$-field itself is that of a classical scalar wave-equation in curved space-time:
\begin{equation} \label{4}
L_K  =  - \frac{{c^4 }} {{32\pi G}}\int {d^4 x\sqrt { - g} } g^{ab} \partial _a K\partial _b K
\end{equation}
where, from equation (\ref{PVLineElement}), in Cartesian coordinates $g^{ab}= \diagonal{\{K,-1/K,-1/K,-1/K\}}$. Evidently this action
is not invariant under arbitrary coordinate transformations, and therefore, unlike GR, the background coordinates here contain some
physical significance.

Several obvious changes and generalisations of equation (\ref{2}) are required to render PV compatible with the Cosmological
Principle:
\begin{enumerate}
\item Long-range EM forces are negligible compared to gravitational forces, whereupon the EM interaction part of the action due to any
charge can be safely ignored.

\item The single charged particle must be replaced by a uniform distribution of neutral mass that is static in the background frame,
whence $v = 0$.

\item $m_0 c^2 \delta ^3 \left( {{\mathbf{r}} - {\mathbf{r}}\left( t \right)} \right) \to \rho_m$ where $\rho_m$ is a uniform (in the
background frame) and constant energy density of matter. Subsequently we will choose $K = 1$ at the present time, so that $\rho_m$
will be the present energy density of matter.

\item On the cosmological scale, the field $K$ is assumed to be isotropic, homogeneous, and therefore a function of time only, for
which it will be convenient to make the substitution $K = a^2 \left( t \right)$, which amounts to the selection of the flat-space
metric out of the three FLRW space-times. The background to this choice is discussed further below.

\item In anticipation of the need to accommodate accelerated expansion, as in conventional GR the action will be supplemented with a
cosmological (henceforth `vacuum') term
\begin{equation} \label{5}
L_{vac}  =  - \rho _v \int {d^4 x\,a^2 }
\end{equation}
where $\rho_v$ has the units of energy per unit volume (the energy density of the vacuum when $a = 1$).
\end{enumerate}

With these changes the PV action, (\ref{2}), becomes
\begin{equation} \label{6}
\begin{split}
L   =  \int {d^4 x} \Biggl\{\frac{{c^2 }} {8\pi G}\left( {a\frac{{da}} {{dt}}} \right)^2  &- \rho _v a^2  - \frac{{\rho_m }} {a}
    -\frac{1} {2}\left[ {\frac{{B^2 }} {{a^2 \mu_0 }} - a^2 \varepsilon_0 E^2 } \right] \Biggr\}
\end{split}\,,
\end{equation}
and consistent with equation (\ref{PVLineElement}) the line element is
\begin{equation} \label{7}
ds^2  = c^2 dt^2 / a^2 \left( t \right) - a^2 \left( t \right)d{\mathbf{x}}^2 \,.
\end{equation}
%
\subsection{Hamiltonian}
%
Carrying out the space integrations, the Lagrangian is
\begin{equation} \label{8}
L = V\left\{ {\frac{{a^2 c^2 }} {{8\pi G}}\left( {\frac{{da}} {{dt}}} \right)^2  - \rho_v a^2  - \frac{{\rho_m }} {a} +
L_{EM} } \right\}
\end{equation}
where the electromagnetic part is
\begin{equation} \label{9}
L_{EM}  = \frac{1} {{2V}}\int {d^3 x} \left\{ {a^2 \varepsilon_0 E^2  - \frac{{B^2 }} {{a^2 \mu_0 }}} \right\} \,.
\end{equation}
The corresponding contributions to the Hamiltonian are
\begin{equation} \label{10}
H = \dot a\frac{{\partial L}} {{\partial \dot a}} - L = \frac{{a^2 c^2 }} {{8\pi G}}\left( {\frac{{da}} {{dt}}} \right)^2 + \rho_v a^2
+ \frac{{\rho_m }} {a} + H_{EM} \left( a \right)
\end{equation}
where $H_{EM} \left( a \right)$ is the electromagnetic energy density. $H_{EM} \left( a \right)$ has the usual (i.e. GR) dependence on
the scale factor $ \rho_r/a^2 $ say, as may be deduced by standard arguments \citep{MTW}. With this in equation (\ref{10}), the total
energy density, $\rho_\Sigma $ say, is
\begin{equation} \label{11}
\frac{{a^2 c^2 }} {{8\pi G}}\left( {\frac{{da}} {{dt}}} \right)^2 + \rho_v a^2  + \frac{{\rho_m }} {a} + \frac{{\rho_r }} {{a^2 }} =
\rho_\Sigma \,.
\end{equation}
%
\subsection{Relation to the GR Friedmann equation}
%
In accord with equation (\ref{7}) the proper time $\tau$ is given by $ d\tau  = dt/a\left( t \right)$, whereupon one recovers the FLRW
line element
\begin{equation} \label{12}
ds^2  = c^2 d\tau ^2  - a^2 \left( \tau  \right)d{\mathbf{x}}^2 \,.
\end{equation}
With this, equation (\ref{11}) written in proper time is the familiar Friedmann equation of GR except for a missing factor of 3 and
some changes in sign. The two equations can be united in the form
\begin{equation} \label{13}
\lambda \frac{{3c^2 }} {{8\pi G}}\left( {\frac{{da}} {{d\tau }}} \right)^2  = \rho_v a^2  + \frac{{\rho_m }} {a} + \frac{{\rho_r }}
{{a^2 }} - \rho_\Sigma
\end{equation}
where in GR $\lambda = 1$, and in PV $\lambda = -1/3$. $\rho_\Sigma $ has been defined so that when $ \rho_v  = 0 $ in GR, a positive
value gives rise to collapse, I.E. it has the sign of `$k$' in the traditional notation. In PV a positive $ \rho_\Sigma $ connotes a
positive net overall energy. In both cases - GR and PV - the magnitude and sign of both $ \rho_\Sigma $ and $ \rho_v $ can be decided,
for example, by regression of the data for $a$ versus $\tau$ onto the corresponding Friedmann equation (i.e. equation (\ref{13}) for
either $\lambda$). Both the mass energy density and the electromagnetic energy density could, in principle, be determined locally,
without reference to either the Friedmann equation or the theoretical framework in which it is to be employed (GR or PV). In any case,
the sign of both will be positive.

Carrying out the usual normalisation using the present value of the Hubble constant
\begin{equation} \label{14}
H_0  \equiv \left. {\left( {\frac{1} {a}\frac{{da}} {{d\tau }}} \right)} \right|_{\tau  = \tau_0 }
    = \frac{da\left({\tau_0 } \right)} {d\tau_0}
\end{equation}
(the scale factor at the present time is 1), equation (\ref{11}) can be written
\begin{equation} \label{15}
\frac{\lambda } {{H_0^2 }}\left( {\frac{{da}} {{d\tau }}} \right)^2  = a^2 \Omega_v  + \frac{{\Omega_m }} {a} + \frac{{\Omega_r }}
{{a^2 }} - \Omega_\Sigma \,,
\end{equation}
where
\begin{equation} \label{16}
\Omega_i  = \frac{{8\pi G\rho_i }} {{3c^2 H_0 ^2 }};\quad i \in \left\{ {m,r,v,\Sigma } \right\}
\end{equation}
and
\begin{equation} \label{17}
\Omega_v  + \Omega_m  + \Omega_r  - \Omega_\Sigma   = \lambda
\end{equation}
are true at all times. Of course equation (\ref{17}) can be used to eliminate any one of the 4 $\Omega$'s in equation (\ref{15}),
restoring it to an equation that is explicitly dependent on just 4 observationally determined constants (3 $\Omega$'s and $ H_0 $ ).
With the additional definition
\begin{equation} \label{18}
\Omega  \equiv \Omega_v  + \Omega_m  + \Omega_r
\end{equation}
employed in the (GR) literature, one has (exclusively in GR) $ \Omega_\Sigma   = \Omega  - 1 $, whereupon $ \Omega  = 1 $ corresponds
to flat space. But $\Omega  = 1 $ with $\Omega$ defined in equation (\ref{18}) has no such special significance in PV, and therefore
the substitution (\ref{18}) is less useful when PV and GR are being compared in parallel, and so will not be employed here. Also, in
the flat-space version of PV presently under consideration, $ \Omega_\Sigma   \ne 0 $ signifies non-zero overall energy, with no
contribution from the spatial variation of $K$. By contrast, in the Friedmann equation of GR, $ \Omega_\Sigma   \ne 0 $ always
signifies a contribution from the (non-flat) geometry - it is not an arbitrary constant of integration.
%
\section{Cosmology}
\subsection{Initial condition}
%
From equation (\ref{15}) the current age as a function of the $\Omega$ is
\begin{equation} \label{19}
\tau  = \frac{1} {H_0 } \int\limits_{a\left( 0 \right)}^1
    {
    da \sqrt
        {
        \frac{\lambda}
            {
            a^2 \Omega_v  + \Omega_m / a + \Omega_r / a^2  - \Omega_\Sigma
            }
        }
    }\,.
\end{equation}
In the case of GR ($\lambda = 1$) the operand of the square root is always positive as $ a \to 0 $ because $ \Omega_r $ is positive.
Therefore in GR one is free to suppose that $ a\left( 0 \right) = 0 $ was the initial state of affairs, and take the age as the proper
time elapsed since then. By contrast in PV ($\lambda = -1/3$) the scale factor cannot shrink to zero else the operand of the square
root will go negative. Defining the age as the proper time since the most recent minimum (including the case that the scale factor
oscillates), $ a\left( 0 \right)$ is the smallest real positive root of
\begin{equation} \label{20}
 a^2\left( 0 \right) \Omega_v  + \Omega_m / {a\left( 0 \right)} + \Omega_r / {a^2\left( 0 \right) } - \Omega_\Sigma   = 0 \,.
\end{equation}
Evidently $ da / d\tau = 0 $ at that point, and therefore the expansion is initially exclusively quadratic - i.e. is not a `bang'.
%
\subsection{Very early history}
%
The early history is decided by the magnitude of the initial (and therefore minimum) scale factor given by the solution of equation
(\ref{20}), for which some estimate of the value of the $\Omega$ coefficients is necessary. The CMB mass-equivalent energy density is
$ 2 \times 10^{ - 31} \text{ kg/m}^ 3 $ (the contribution from stellar radiation is around 20\% of this value and can be ignored
relative to the imprecision of this analysis). Taking $ H_0  \approx 70 \text{ km/s/Mpc} $, giving a `Hubble time' $ 1/H_0 \approx 14$
Gyr, gives (from equation (\ref{16})) $ \Omega_r \approx 2 \times 10^{-5} $. The widest possible range for the matter density is $
0.005 < \Omega_m < 0.3 $ (absorbing therein any uncertainty in $H_0$). The lower limit corresponds to luminous baryonic matter (only).
The upper limit is supplied by, for instance, fitting super-nova data to the (GR) Friedmann equation - see for example
\citet{Perlmutter} and \citet{Riess}. Although the applicability to PV of such a large value for the upper limit is questionable, it
will turn out not to make any difference to the conclusion of this section.

To achieve a lepton era requires a temperature of around $T_{lepton}  \approx 10^{10} \text{ K} $ and therefore an initial scale
factor of no more than $a\left( 0 \right) = T_{CMB}/ T_{lepton}  \sim 3 \times 10^{ - 10}$. At those temperatures, the matter term in
equation (\ref{20}) is certainly negligible compared to the radiation term, and therefore one can take
\begin{equation} \label{21}
a^2\left( 0 \right) \Omega_v  + \Omega_r/a^2\left( 0 \right) - \Omega_\Sigma   = 0 \,.
\end{equation}
Equation (\ref{17}) gives that the magnitudes of $ \Omega_v $ and $ \Omega_\Sigma $ can differ at most by a number of order unity,
from which it is deduced that the first term in equation (\ref{21}) is negligible compared with the second term. Therefore $
\Omega_\Sigma \approx \Omega_r / a\left( 0 \right)^2 \approx 2 \times 10^{14}.$ But such a large number has un-physical consequences
on the age, for example, which can be shown as follows. With reference to equation (\ref{19}), over the interval $ a\left( 0 \right)
\leqslant a \leqslant 1 $, the very large $ \Omega_r $ makes the square root very small everywhere except when $a$ is very close to $
a\left( 0 \right) $. Since $ \Omega_v $ and $ \Omega_\Sigma $ have similar magnitudes, it follows that $ a^2 \left| {\Omega_v }
\right| \ll \left| {\Omega_\Sigma } \right| $ and $ a\Omega_m  \ll \Omega_r $ over the region of importance. Therefore equation
(\ref{19}) is well approximated by
\begin{equation} \label{22}
\tau  \approx  \frac{1} {{\sqrt 3 H_0 }}\int\limits_{a\left( 0 \right)}^1 {da\frac{1} {{\sqrt {\Omega_\Sigma   - {{\Omega_r }
\mathord{\left/
 {\vphantom {{\Omega_r } {a^2 }}} \right.
 \kern-\nulldelimiterspace} {a^2 }}} }}} \,.
\end{equation}
Rearranging a little and then performing the integral:
\begin{equation} \label{23}
\begin{split}
\tau    &\approx  \frac{1} {{\sqrt {3\Omega_r } H_0 }}\int\limits_{a\left( 0 \right)}^1 {da\frac{a} {{\sqrt {\left( {{a
            \mathord{\left/ {\vphantom {a {a\left( 0 \right)}}} \right.
            \kern-\nulldelimiterspace} {a\left( 0 \right)}}} \right)^2  - 1} }}}  \\
        &=  \left. {\frac{{{a^2\left( 0 \right)} \sqrt {\left( {{a \mathord{\left/
            {\vphantom {a {a\left( 0 \right)}}} \right.
            \kern-\nulldelimiterspace} {a\left( 0 \right)}}} \right)^2  - 1} }}
            {{\sqrt {3\Omega_r } H_0 }}} \right|_{a\left( 0 \right)}^1
        \approx  \frac{{a\left( 0 \right)}} {{\sqrt {3\Omega_r } H_0 }} \,.
\end{split}
\end{equation}
Using the numbers above, this gives an age of only 540 yrs. Consequently it must be concluded that PV cosmology is
incapable of accommodating either a lepton or a baryon era.
%
\subsection{Radiation era}
%
Existence of a radiation era requires that the initial radiation energy density is at least equal to that of matter: ${a\left( 0
\right) = {{\Omega_r } \mathord{\left/
 {\vphantom {{\Omega_r } {\Omega_m }}} \right.
 \kern-\nulldelimiterspace} {\Omega_m }}\,.}$ Using this in equation (\ref{20}) (where $\Omega_v $
comes from equation (\ref{17}), but in any case $a^2\left( 0 \right) \left| {\Omega_v } \right| \ll \left| {\Omega_\Sigma  } \right|$
and so can be neglected) gives
\begin{equation} \label{23.5}
\Omega_\Sigma   \approx {{2\Omega_r } \mathord{\left/
 {\vphantom {{2\Omega_r } {a\left( 0 \right)}}} \right.
 \kern-\nulldelimiterspace} {a^2\left( 0 \right)}}  = {{2\Omega_m^2 } \mathord{\left/
 {\vphantom {{2\Omega_m^2 } {\Omega_r }}} \right.
 \kern-\nulldelimiterspace} {\Omega_r }} \leqslant 2.5 \,,
\end{equation}
the upper limit corresponding to $ \Omega_m  = 0.005. $ The approximation used for the lepton era, (\ref{23}), gives
\begin{equation} \label{24}
\tau  \approx  \frac{{a\left( 0 \right)}} {{\sqrt {3\Omega_r } H_0 }}
    \approx  \sqrt {\frac{{\Omega_r }} {3}} \frac{1} {{\Omega_m H_0}}
\end{equation}
which, using the numbers above, gives an age less than 7.2 Gyr. However, that approximation required $\Omega_\Sigma   \gg 1$, so this
result cannot be expected to be very accurate. In fact an accurate numerical evaluation of equation (\ref{19}) with $ \Omega_\Sigma =
2.5 $ gives an age of 6.6 Gyr or less, the upper limit corresponding to $ \Omega_m  = 0.005 $. This age is unacceptably short - too
small to accommodate main-sequence behaviour in globular clusters for example \citep{Peacock}. It must be concluded that PV cosmology
cannot encompass an era of radiation domination. In short, combining the conclusions of this and the previous section, PV cosmology
starts with a big whimper.
%
\subsection{Mass density}
%
Current estimates of the matter density based upon observation of the proportions of the light elements inferred from big-bang
nucleosynthesis (BBN) are around $ \Omega_m \approx 0.1 $. But given the absence of a baryonic era in PV, the theory is denied the
chain of inference that leads from BBN to this value. Further, though recent observations of super-nova lead to $ \Omega_v \approx
0.7,\quad \Omega_m  \approx 0.3,\quad \Rightarrow \Omega_\Sigma   \approx 0 $ - the difference between this value of $\Omega_m$ and
the value inferred from BBN is missing matter - these super-nova $\Omega_i$ are due (effectively) to regression of the data onto the
curve predicted by equation (\ref{15}) when $ \lambda = 1$, which results cannot be carried over into PV, where $ \lambda  =  - {1
\mathord{\left/ {\vphantom {1 3}} \right. \kern-\nulldelimiterspace} 3}. $ With the bases for large contributions from dark and
missing matter gone, it seems that at least for the present PV should be examined with an $ \Omega_m $ much closer to the observed -
luminous - density, plus whatever component is found to come from the neutrino. This conclusion is vulnerable to revision however, if
a significant dark-matter contribution turns out to be demanded for other reasons, such as, for example, anomalous rotation of spiral
galaxies.
%
\subsection{Cosmological red shift}
%
A consequence of the non-zero initial scale factor is the upper limit it imposes on the cosmological redshift of distant objects:
$\max \left( z \right) = 1/a\left( 0 \right) - 1\,,$ where $a\left( 0 \right) $ is the smallest positive root of equation (\ref{20})
($ a\left( 0 \right) $ - and therefore $\max \left( z \right)$ - are functions of the $\Omega_i$). Fig. 1 is a plot of the maximum
redshift versus matter density for $ \Omega_\Sigma   = 0 $ and $ \Omega_\Sigma   =  \pm 0.1 $ obtained from a numerical solution of
equation (\ref{20}). It is observed that PV can accommodate all the observed redshifts provided there is an overall deficit in the
energy density. At the time of writing the largest claimed redshift for a radio galaxy is TN J0924-2201 at $z = 5.19$ \citep{Breugel}.
But Fig. 2 shows that if, axiomatically, (say) $ \Omega_\Sigma   = 0 $, then PV is incompatible with a cosmological origin of
redshifts beyond 3, since these correspond to $ \Omega _m  < 0.005 $ - the luminous mass density. Since the observational upper limit
on the redshift is subject to constant review, it is not possible to use that to deduce a firm relation between $ \Omega_\Sigma $ and
$\Omega_m $ in PV. Loose constraints on the range of these parameters can be inferred from Figs. 2 and 3, which shows simultaneously
how they affect the maximum redshift, the current age, and the deceleration parameter over different ranges. In both figures redshifts
beyond 10 are increasingly crowded and have been suppressed. Broadly, one sees PV requires that
\begin{equation} \label{25}
0.1 < \Omega_\Sigma   < 0.4,\quad 0.005 < \Omega_m  < 0.1
\end{equation}
the intersection of these two domains giving a reasonable range of ages, and a range of redshifts that encompass the most distant
observations to date. Note however that in light of the arguments of the previous section (absence of motivation for dark or missing
matter), current observations constrain $\Omega_m$ to the lower end of this range.
%
\begin{figure}
\label{fig1}%
\plotone{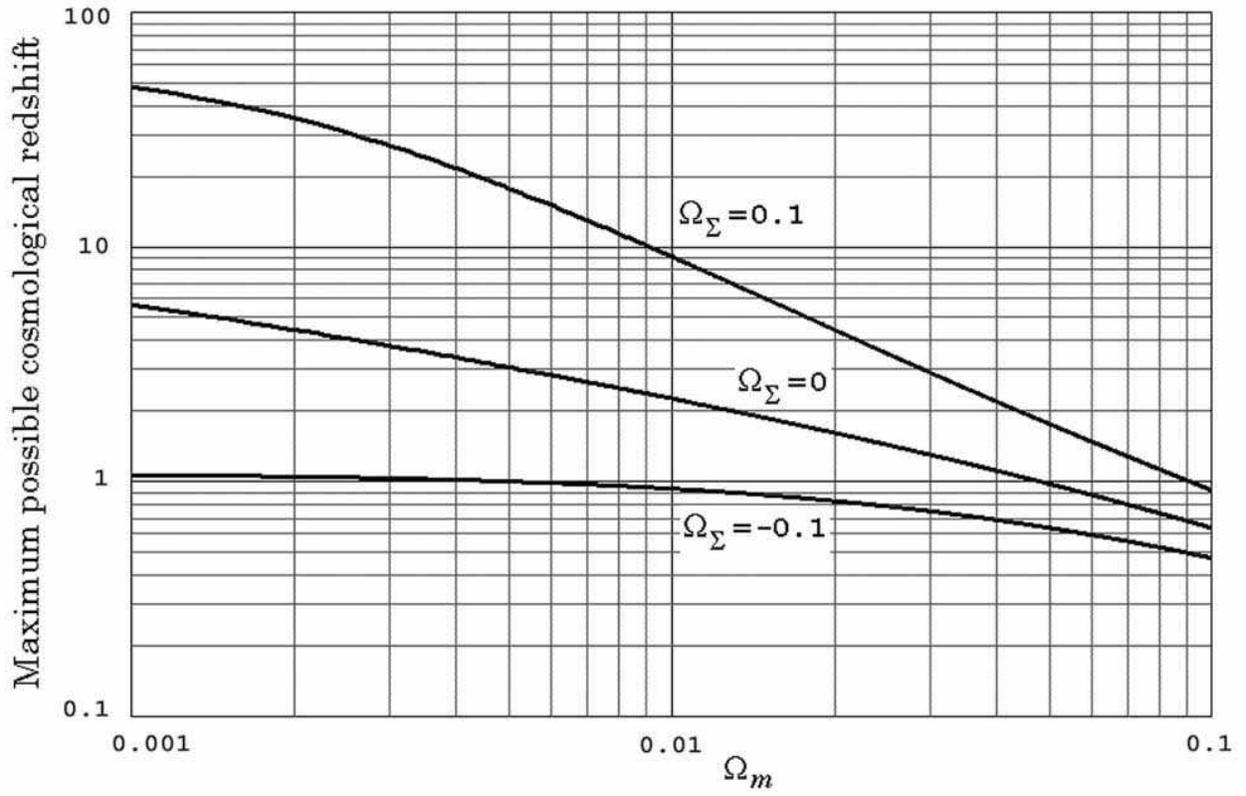} 
    \caption{Maximum currently visible redshift versus the matter density $\Omega_m$ when $1/H_0 = 14$ Gyr.}
\end{figure}
%
%
\subsection{Deceleration parameter}
%
It would appear from the form of equation (\ref{15}) that the deceleration parameter $ q \equiv  - a\ddot a / \dot a^2 $, since it is
independent of the scale of both $a$ and $\tau$, is independent of $\lambda$. However, taking into account that the four $\Omega_i$ on
the right hand side of equation (\ref{15}) are internally related in a way that depends on $\lambda$ through equation (\ref{17}), the
PV prediction for the deceleration parameter as a function of the three remaining true degrees of freedom differs from that of GR.
Differentiating both sides of equation (\ref{15}) one obtains
\begin{equation} \label{26}
\frac{{2\lambda }} {{H_0^2 }}\frac{{da}} {{d\tau }}\frac{{d^2 a}} {{d\tau ^2 }} = \frac{{da}} {{d\tau }}\left( {2a\Omega_v  -
\frac{{\Omega_m }} {{a^2 }} - \frac{{2\Omega_{EM} }} {{a^3 }}} \right)
\end{equation}
and therefore
\begin{equation} \label{27}
\begin{split}
q \equiv  - \frac{{a\ddot a}} {{\dot a^2 }}
    &=  - \frac{a\left( a\Omega_v  - \Omega_m  / 2a^2 - \Omega_r / a^3 \right)}
        {a^2 \Omega_v  + \Omega_m / a + \Omega_r/a^2 - \Omega_\Sigma} \\
    &=  \frac{a\Omega_m /2+ \Omega_r  - a^4 \Omega_v }
        {a^4 \Omega_v  - a^2 \Omega_\Sigma   + a\Omega_m  + \Omega_r } \,.
\end{split}
\end{equation}
It may be observed that in PV, because the first derivative of the scale factor is zero when $\tau = 0$ (it is infinite in the FLRW
cosmologies), the deceleration parameter defined above is negative infinite there. The present (and therefore `local') value for both
theories is
\begin{equation} \label{28}
\begin{split}
 q_0
    & \equiv \left. q \right|_{a = 1}  =  \frac{\Omega_m /2 + \Omega_r  - \Omega_v }
        {\Omega_v  - \Omega_\Sigma   + \Omega_m  + \Omega_r }\\
    &=  \frac{1} {\lambda }\left( \Omega_m /2 + \Omega_r  - \Omega_v \right)
    =  \frac{1}{\lambda }\left( 3\Omega_m/2 + 2\Omega_r  - \Omega_\Sigma  \right) - 1
\end{split}
\end{equation}
where equation (\ref{17}) was used. For PV in particular
\begin{equation} \label{29}
\begin{split}
q_0  &=   - 3\left( \Omega_m /2 + \Omega_r  - \Omega_v \right) \\
     &=  - 3\left( 3\Omega_m /2 + 2\Omega_r  - \Omega_\Sigma  \right) - 1\,.
\end{split}
\end{equation}
Rearranging equation (\ref{29}), one sees that the condition for a locally accelerating universe ($ q_0  < 0 $ ) is that
\begin{equation} \label{30}
\Omega_\Sigma   < 3\Omega_m /2 + 2\Omega_r  + 1/3\,.
\end{equation}
From equation (\ref{25}) it is observed that the plausible range of $ \Omega_\Sigma $ and $ \Omega_m $ encompasses both local
acceleration and deceleration. Specifically, the range of $q_0$ consistent with equation (\ref{25}) is
\begin{equation} \label{31}
 - 1.15 \leqslant q_0  \leqslant 0.18\,.
\end{equation}
Contours of constant $q_0$ versus $ \Omega_\Sigma $ and $ \Omega_m $ are plotted in Figs. 2 and 3.

Some caution is necessary in comparing published results of $q_0$ with the value predicted by equation (\ref{29}): where $q_0$ is
deduced effectively from a parametric fit of the Friedmann equation to the scale factor as deduced from redshift and brightness data,
e.g. \citet{Riess}, the reported value will be applicable exclusively to GR and not PV.
%
%
\begin{figure*}
\label{fig2}
\plotone{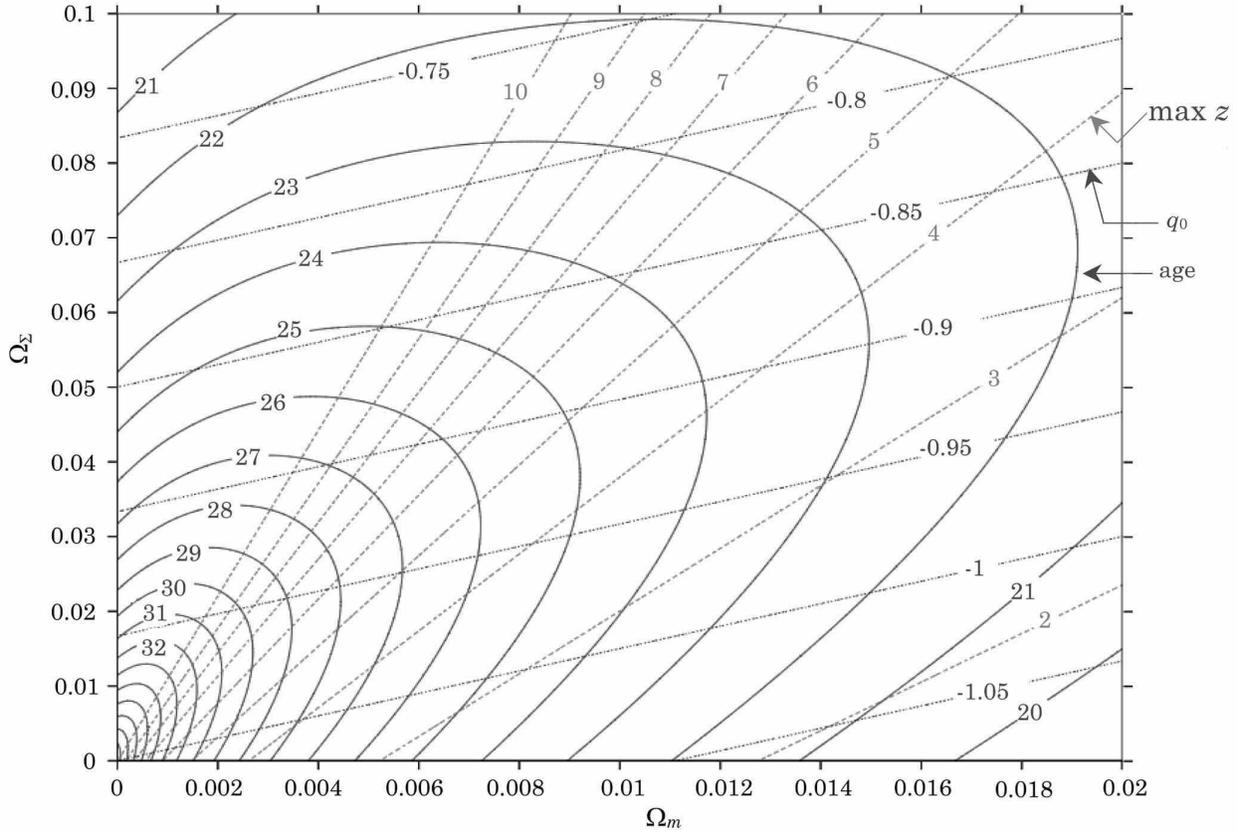} 
    \caption{For values of $\Omega_m$ close to the density of luminous matter,
    contours showing the dependency on $\Omega_{\Sigma}$ and $\Omega_m$ of: the current age in Gyr
(solid ellipse-like contours), the maximum possible cosmological redshift (broken lines), and the current value of the deceleration
parameter $q_0$ (dotted lines). Redshifts beyond 10 are not shown.}
\end{figure*}
\begin{figure*}
\label{fig3}
\plotone{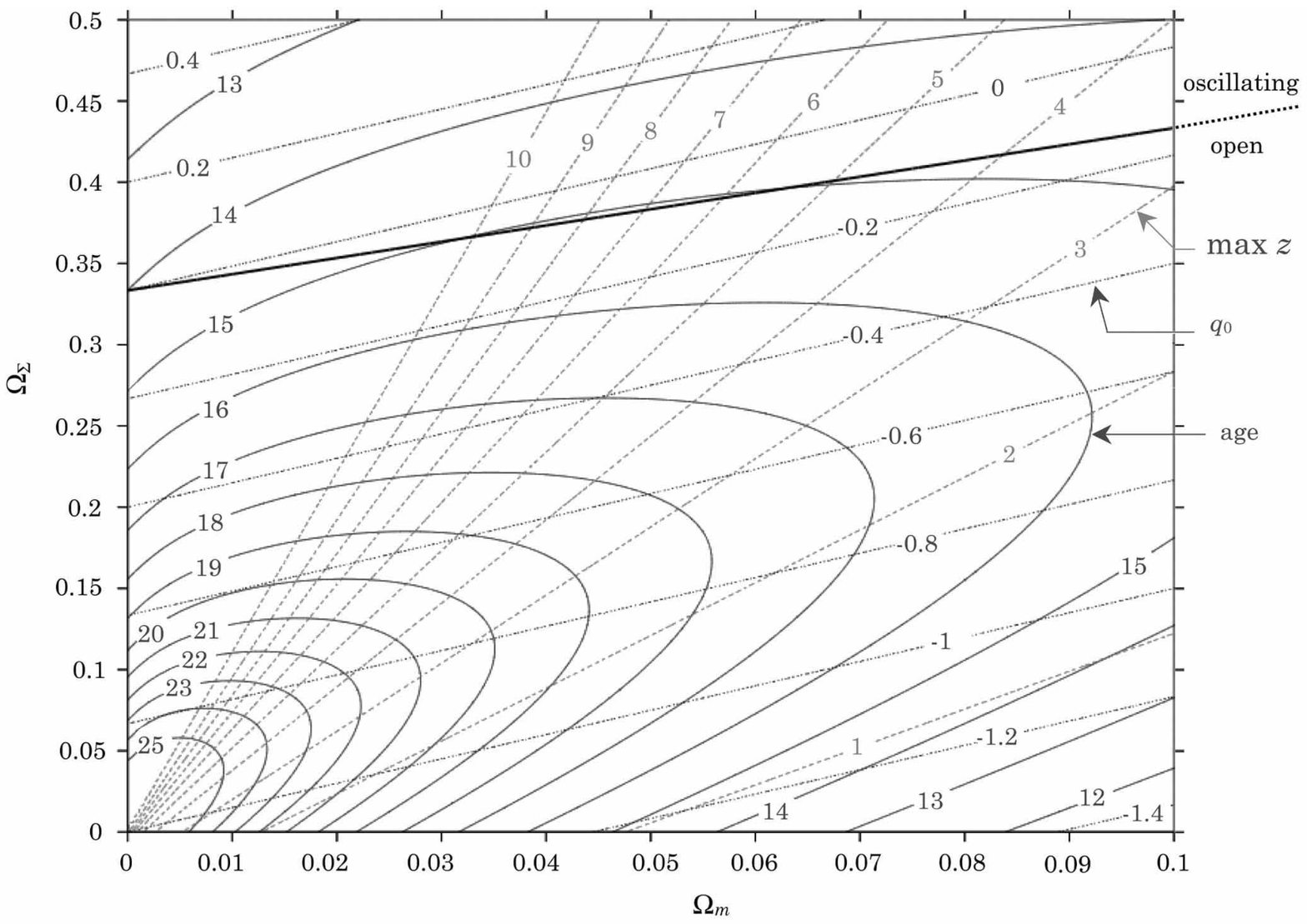} 
    \caption{For a large range of $\Omega_m$, contours showing the dependency on $\Omega_{\Sigma}$ and $\Omega_m$ of:
    the current age in Gyr (solid ellipse-like contours), the maximum possible cosmological redshift (broken lines),
    the current value of the deceleration parameter $q_0$ (dotted lines),
    and the boundary between oscillating and open cosmologies (heavy line). Redshifts beyond 10 and ages beyond 25 Gyr are not shown.}
\end{figure*}
%
\subsection{Future collapse and oscillation}
%
The condition for future collapse is that the right hand side of equation (\ref{15}) becomes zero at some $ a > 1 $, i.e. equation
(\ref{20}) possesses more than one real positive solution. At that point $\left( da /d\tau  \right)^2  = 0$ and the expansion stalls
and will subsequently reverse. Combining equations (\ref{17}) and (\ref{20}) gives that future collapse requires
\begin{equation} \label{32}
a^2 \Omega_v  - 1/3 - \Omega_v  - \Omega_m  - \Omega_r  + \frac{{\Omega_m }} {a} + \frac{{\Omega_r }} {{a^2 }} = 0
\end{equation}
to have a solution in the range $a > 1$. Solving equation (\ref{32}) for $ \Omega_v $,
\begin{equation} \label{33}
\Omega_v  = \frac{1} {{3\left( {a^2  - 1} \right)}} + \frac{{\Omega_m }} {{a\left( {a + 1} \right)}} + \frac{{\Omega_r }} {{a^2 }}
\end{equation}
one sees that the range $ 1 < a < \infty $ corresponds to
\begin{equation} \label{34}
0 < \Omega_v  < \infty\,.
\end{equation}
An infinite vacuum density corresponds to a turn-around point at the present time when $a = 1$. (A turn-around earlier than this has
effectively been excluded by the normalisation process in which the $\Omega$ defined in equation (\ref{16}) are based upon on $a = 1$
present-day values.) The condition $ \Omega_v  = 0 $ is a special `critical' case and is discussed separately below. Inserting
equation (\ref{34}) into equation (\ref{17}), the constraint on the total energy for future collapse is
\begin{equation} \label{35}
1/3 + \Omega_m  + \Omega_r  < \Omega_\Sigma   < \infty\,.
\end{equation}
Solutions of equation (\ref{15}) satisfying equation (\ref{35}) are necessarily oscillating and symmetric about the times of maximum
and minimum scale factor. If the constraint equation (\ref{35}) is not satisfied, then the expansion is monotonic (`open'). The locus
of this constraint is shown as a heavy line in Fig. 3. (The line may be regarded as a contour of constant period for the particular
case that the period is infinite - other equal-period contours could be drawn above this line.) It may be observed that a limited
range of plausible combinations of age and maximum redshift lie above this line - the majority of plausible combinations lie below.
Comparison of equation (\ref{35}) with equation (\ref{30}) reveals that monotonic expansion is a sufficient but not necessary
condition for local acceleration. (Consequently, a local acceleration does not guarantee continued expansion.)

As an example, PV predicts an oscillating cosmology using values $ \Omega_m  = 0.01 $ and $ \Omega_\Sigma = 0.4 $, and therefore -
from equation (\ref{17}) - $ \Omega_v  = 0.06 $. The first positive root of equation (\ref{20}) is $ a\left( 0 \right) = 0.027 $,
implying a maximum observable redshift of 36. Using this value for the lower limit of the integration in equation (\ref{19}), and
using $1/H_0 = 14$ Gyr, one obtains a present age of 14 Gyr. From equation (\ref{29}), the present deceleration parameter is 0.16.
From further numerical investigation it is deduced that for these values the expansion started to decelerate at an age of 7 Gyr, i.e.
7 Gyr ago. Figs. 4 and 5 show the evolution of the scale factor over time from numerical solution of equation (\ref{15}) for these
particular values. The oscillation has a period of 109 Gyr and a maximum scale-factor of 2.65. Also included for comparison is a plot
of the scale factor according to GR for the currently favoured values of $ \Omega_m  = 0.3 $ and $ \Omega_\Sigma   = 0 $, whereupon
equation (\ref{17}) gives that $ \Omega_v  = 0.7 $.
%
\subsection{Monotonic expansion}
%
Any combination of $ \Omega _\Sigma $ and $ \Omega_m $ below the heavy line in Fig. 3 gives rise to monotonic expansion. If the mass
density is left unchanged from the previous example at $ \Omega_m  = 0.01 $, and $ \Omega_\Sigma $ is reduced by 0.1 from $
\Omega_\Sigma   = 0.4 $ to $ \Omega_\Sigma   = 0.3 $, then, from equation (\ref{17}), $ \Omega_v =  - 0.04 $. Recalling that the
condition for oscillation is that the vacuum density is positive, this combination must lead to an open universe. The first positive
root of equation (\ref{20}) is $ a\left( 0 \right) = 0.035 $, implying a maximum observable redshift of 28. Using this value for the
lower limit of the integration in equation (\ref{19}), and using $ 1/H_0 = 14 $ Gyr, one obtains a present age of 15 Gyr. From
equation (\ref{29}), the present deceleration parameter is -0.15 - i.e. a positive acceleration.
%
\begin{figure}
\label{fig4}
\plotone{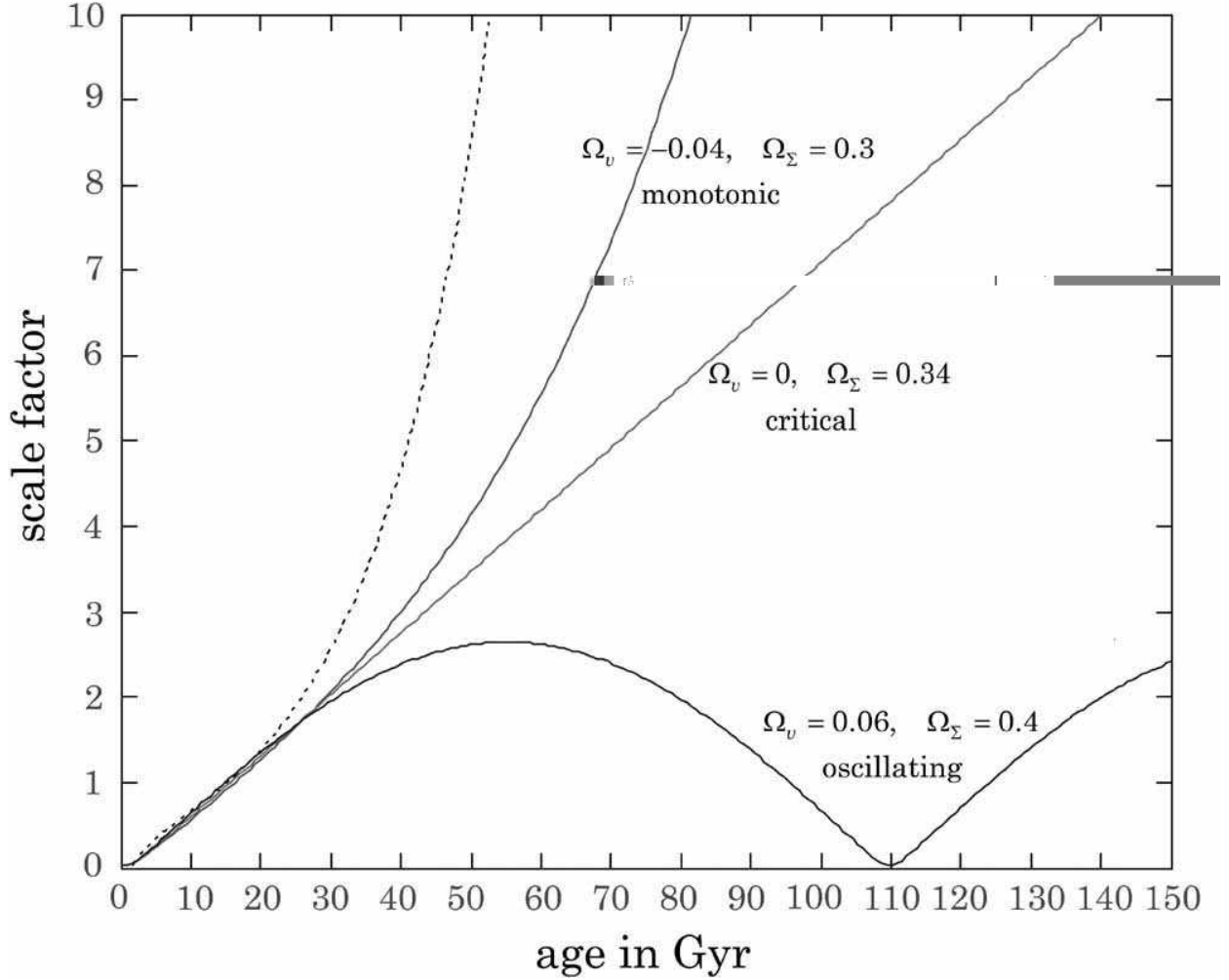} 
    \caption{Evolution of scale factor over 150 Gyr. The solid lines are predictions of PV ($\lambda = -1/3$) given $1/H_0  = 14$
    Gyr and assuming a mass density close to the visible value, $ \Omega_m = 0.01 $, for three different values of vacuum energy density,
    and consequent value for $ \Omega_\Sigma $. The dotted line is the prediction of GR ($ \lambda = 1 $) for the currently favoured
    values $ \Omega_m = 0.3,\quad \Omega_v = 0.7,\quad \Omega_\Sigma = 0\,. $}
\end{figure}
\begin{figure}
\label{fig5}
\plotone{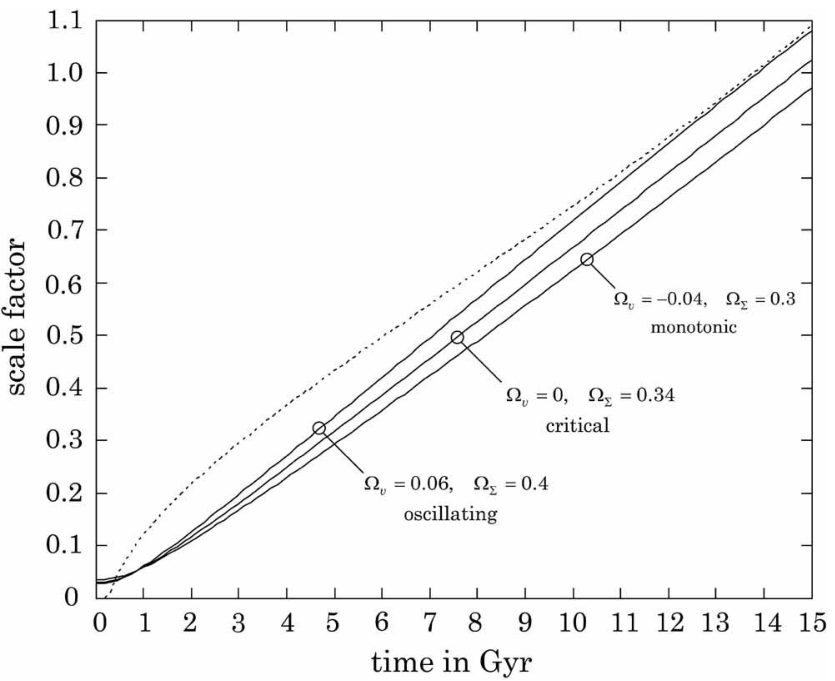} 
    \caption{Evolution of scale factor to 15 Gyr for the same set of parameters as in Fig. 4.
    The solid lines are predictions of PV ($ \lambda = -1/3$) given $1/H_0  = 14$
    Gyr and assuming a mass density close to the visible value, $\Omega_m = 0.01$,
    for three different values of vacuum energy density, and consequent value for $ \Omega_\Sigma $. The dotted
    line is the prediction of GR ($ \lambda = 1 $) for the currently favoured values
    $\Omega_m = 0.3,\quad \Omega_v = 0.7,\quad \Omega_\Sigma = 0\,.$}
\end{figure}
%
%
\subsection{Critical expansion}
%
If there is no contribution from the vacuum at all, then from equation (\ref{33}) $ \Omega_v  = 0 \Rightarrow a = \infty\,,$ i.e. PV
predicts an oscillating universe with an infinite maximum scale factor and therefore with an infinite period. Constructing an example
using the same mass-density as in the previous example, $ \Omega_m  = 0.01\,,$ equation (\ref{17}) gives that $ \Omega_\Sigma   =
0.34\,.$ The first positive root of equation (\ref{20}) is $ a\left( 0 \right) = 0.031\,,$ implying a maximum observable redshift of
31. Using this value for the lower limit of the integration in (19), and using $1/H_0 = 14$ Gyr, one obtains a present age of close to
14.5 Gyr. From equation (\ref{29}), the present deceleration parameter is -0.015 - i.e. a small positive acceleration. The scale
factor versus time for this configuration is given in Figs. 4 and 5 along with the previous open and oscillating universe
configurations, and the currently favoured GR evolution.
%
\subsection{Comparisons}
%
For the particular values considered above, there is little difference at the present time between the oscillating, monotonic, and
critical expansions, and the currently favoured GR prediction. The present-day values and slopes are within a few percent of each
other, and the second derivative of the scale factor (and therefore $q_0$) is close to zero for all four trajectories. It is not
significant that, of the three PV scenarios considered, the oscillating evolution is the closest to the GR prediction, because there
is sufficient leeway in the mass and vacuum density parameters to change the ranking. But it is noteworthy that with little or no
vacuum term, and no clear need for `missing' mass, PV comes close to the favoured GR-determined trajectory, achieving accord with
current estimates of the Hubble constant, model-independent estimate of age, luminous mass-density, and accommodation of the maximum
observed redshift.
%
\subsection{Summary of cosmological findings}
%
The principal characteristics of the scalar, flat-space, PV cosmology investigated above are summarized below:
\begin{enumerate}
\item Compared with GR, the scale factor evolves according to a Friedmann-like equation in which the $\dot a^2 $ term is multiplied by
$-1/3$, with all the other terms appearing in the same way.

\item A consequence is that PV cosmology has no big bang, but starts smoothly from a non-zero scale factor.

\item The initial scale factor is too large for a baryon, lepton, or radiation era.

\item Consequently PV can explain neither present proportions of light elements nor the cosmic microwave background as the result of
high temperatures in the very early history.

\item The non-zero scale factor gives rise to a maximum possible redshift.

\item PV can predict an age, deceleration parameter, and maximum possible redshift for particular values of mass density and overall
system energy compatible with observation, without missing mass, and without a vacuum term.

\item PV cosmologies compatible with observation can be monotonic, critical, or oscillating.

\item Compatibility demands that PV contain a significant contribution from the vacuum, or an offset to the total energy, or both.
\end{enumerate}
%
\section{Gravitational radiation}
\subsection{Gravitational potential in the far-field}
%
$K$-field radiation induced by the motion of `local' matter can be decoupled from Cosmologically induced variations provided the
(retarded) distance to the radiation source is small on the Cosmological scale. Then, neglecting any EM contribution (including
pressure), and the vacuum term, the appropriate action is
\begin{equation} \label{36}
\begin{split}
L =  - c^2 \int {d^4 x} \Biggl\{  \frac{1} {{32\pi G}} \left( {c^2 \left( {\nabla \log K} \right)^2  - \left( {\frac{{\partial K}}
{{\partial t}}} \right)^2 } \right)
     + \rho_m \sqrt {\frac{1} {K} - \frac{{v^2 }} {{c^2 }}K} \Biggr\}
\end{split}
\end{equation}
where $ \rho_m \left( {{\mathbf{x}},t} \right) $ is presumed known distribution of matter and $ {\mathbf{v}}\left( {{\mathbf{x}},t}
\right) $ its local velocity field. Writing $ K\left( {{\mathbf{x}},t} \right) = 1 + 2\phi \left( {{\mathbf{x}},t} \right) $
 ($\phi$ is the dimensionless Newtonian gravitational potential) and going immediately to the weak-field limit $
\left| \phi  \right| \ll 1 $, the action is
\begin{equation} \label{37}
\begin{split}
L   & =   - c^2 \int {d^4 x} \Biggl\{ \frac{1} {8\pi G} \left( {c^2 \left( {\nabla \phi } \right)^2  - \left(
        {\frac{{\partial \phi }} {{\partial t}}} \right)^2 } \right) \\
    & + \rho_m \Biggl( {\sqrt {1 - {{v^2 } \mathord{\left/
        {\vphantom {{v^2 } {c^2 }}} \right.
        \kern-\nulldelimiterspace} {c^2 }}}
        - \phi {\frac{{1 + {{v^2 } \mathord{\left/ {\vphantom {{v^2 } {c^2 }}} \right.
         \kern-\nulldelimiterspace} {c^2 }}}} {{\sqrt {1 - {{v^2 } \mathord{\left/{\vphantom {{v^2 } {c^2 }}} \right.
        \kern-\nulldelimiterspace} {c^2 }}} }}}  + O\left( {\phi ^2 } \right)} \Biggr) \Biggr\}\,.
\end{split}
\end{equation}
To first order, the Euler equation for $\phi$ is a scalar wave equation sourced by the matter:
\begin{equation} \label{38}
\frac{1} {c^2 }\frac{{\partial ^2 \phi }} {\partial t^2 } - \nabla ^2 \phi  = \frac{4\pi G\rho_m } {c^2 }\,.
\end{equation}
In the case of relativistic speeds, it is easily seen from equation (\ref{37}) that the right hand side of the above has an additional
coefficient $ {{\left( {1 + {{v^2 } \mathord{\left/
 {\vphantom {{v^2 } {c^2 }}} \right.
 \kern-\nulldelimiterspace} {c^2 }}} \right)} \mathord{\left/
 {\vphantom {{\left( {1 + {{v^2 } \mathord{\left/
 {\vphantom {{v^2 } {c^2 }}} \right.
 \kern-\nulldelimiterspace} {c^2 }}} \right)} {\sqrt {1 - {{v^2 } \mathord{\left/
 {\vphantom {{v^2 } {c^2 }}} \right.
 \kern-\nulldelimiterspace} {c^2 }}} }}} \right.
 \kern-\nulldelimiterspace} {\sqrt {1 - {{v^2 } \mathord{\left/
 {\vphantom {{v^2 } {c^2 }}} \right.
 \kern-\nulldelimiterspace} {c^2 }}} }}\,.$
From application of the virial theorem \citet{Dicke2} interprets this term as the total energy. His interpretation rests, however,
upon the assumption of particles interacting principally electromagnetically, whereas the `particle' sources of mass contemplated here
are stars interacting gravitationally. Further, in the latter case relativistic orbital speeds are possible only in the presence of
strong fields - the weak-field limit is necessarily non-relativistic. Consequently equation (\ref{38}) cannot be extended to the
relativistic domain, which obviates interpretation of the relativistic terms and discussion of the applicability of the virial theorem
to this case.

The attention here is exclusively on the power lost to gravitational radiation, for which it is sufficient to take the far-field limit
of the retarded Green function, whereupon the solution of equation (\ref{38}) is
\begin{equation} \label{39}
\phi \left( {{\mathbf{r}},t} \right) \to \phi_{far} \left( {{\mathbf{r}},t} \right) = \frac{G} {{c^2 r}}\int {d^3 {\mathbf{r'}}} \rho
_m \left( {{\mathbf{r'}},t - \left| {{\mathbf{r}} - {\mathbf{r'}}} \right|/c} \right)
\end{equation}
for large $r$. Expanding,
\begin{equation} \label{40}
\begin{split}
\left| {{\mathbf{r}} - {\mathbf{r'}}} \right|
    & = \sqrt {r^2  + {r'}^2  - 2{\mathbf{r}}{\mathbf{.r'}}} \\
    & = r\left( {1 - \frac{{{\mathbf{\hat r}}{\mathbf{.r'}}}} {r} + \frac{{{r'}^2  - \left( {{\mathbf{\hat r}}{\mathbf{.r'}}} \right)^2 }}
        {{2r^2 }} + ...} \right)
\end{split}
\end{equation}
the far-field potential is
\begin{equation} \label{41}
\begin{split}
\phi  _{far}  \left( \mathbf{r},t \right) = \frac{2G} {c^2 r}
     \int {d^3 \mathbf{r'}}
      \Biggl\{ 1 + {\mathbf{\hat r}}{\mathbf{.r'}}\frac{1} {c}\frac{\partial } {{\partial t}}
     + \frac{\left( {\mathbf{\hat r}}{\mathbf{.r'}} \right)^2 } {2c^2 }\frac{\partial ^2 } {\partial t^2 }
        + O\left( \frac{1}{r} \right) \Biggr\} \rho_m \left( \mathbf{r'},t - r/c \right)\,.
\end{split}
\end{equation}
Since in the Laurent expansion of the gravitational potential only terms proportional to $1/r$ contribute to the radiated power, the
higher order terms can be ignored. The first two terms of equation (\ref{41}) are the monopole and dipole contributions to the
potential, proportional respectively to the total energy and momentum of the source matter. For a closed system (ignoring the
radiation itself), these are constant in time and cannot contribute to radiation since the latter is driven by non-vanishing
time-derivatives of the potential (see below). For radiation, therefore, it is sufficient to take
\begin{equation} \label{42}
\phi_{rad} \left( {{\mathbf{r}},t} \right) = \frac{G} {{c^4 r}}\frac{{\partial ^2 }} {{\partial t^2 }}\int {d^3 {\mathbf{r'}}} \left(
{{\mathbf{\hat r}}{\mathbf{.r'}}} \right)^2 \rho_m \left( {{\mathbf{r'}},t - r/c} \right)\,.
\end{equation}
%
\subsection{Radiation and radiated power}
%
The Euler equation for distant test particles gives a force proportional to $ \nabla \phi $ which, in the far field, is directed away
from the source - the radiation is longitudinally polarized. For maximum sensitivity therefore, PV radiation detectors should be
aligned correspondingly. Since this is orthogonal to the direction predicted by GR, direct detection of gravitational radiation from a
Cosmological source would be a decisive test in favor of one of the two theories.

Disregarding the polarization, the total power radiated is the rate of energy crossing a closed surface far from the source.
Corresponding to the Lagrangian density $ \mathcal{L},\;\;L =  \int {d^4 x} \mathcal{L} \,,$ the energy density in the radiation field
is
\begin{equation} \label{43}
\begin{split}
\mathcal{H}\left( {\phi_{rad} } \right)
    & = \frac{{\partial \mathcal{L}}} {{\partial \left( {\partial \phi_{rad} /\partial t}
\right)}}\frac{{\partial \phi_{rad} }} {{\partial t}} - \mathcal{L} \\
    & = \frac{{c^4 }} {{8\pi G}}\left[ {\left( {\nabla \phi_{rad} } \right)^2  + \frac{1} {{c^2 }}\left( {\frac{{\partial \phi_{rad} }}
{{\partial t}}} \right)^2 } \right]\,.
\end{split}
\end{equation}
At sufficiently great distances from the source, the field momentum is exclusively radial and therefore the power $P$, say, is
\begin{equation} \label{44}
\begin{split}
P   & = c\mathop{{\int\!\!\!\!\!\int}\mkern-21mu \bigcirc}
 {d\Omega_{\mathbf{r}} } r^2 \mathcal{H}\left( {\phi_{rad} } \right) \\
    & = \frac{{c^5 }}
{{8\pi G}}\mathop{{\int\!\!\!\!\!\int}\mkern-21mu \bigcirc}
 {d\Omega_{\mathbf{r}} } r^2 \left\{ {\left( {\nabla \phi_{rad} } \right)^2  + \frac{1}
{{c^2 }}\left( {\frac{{\partial \phi_{rad} }} {{\partial t}}} \right)^2 } \right\}\,.
\end{split}
\end{equation}
With reference to equation (\ref{42}), the spatial gradient of $ \phi $ will introduce additional powers of $1/r$ when it operates on
the $ \mathbf{\hat r} $, and therefore does not contribute to the power through a surface at a large distance from the source. The
spatial gradient operating on the retarded time \emph{does} contribute however:
\begin{equation} \label{45}
\left( {\nabla f\left( {t - r/c} \right)} \right)^2  = \frac{1} {{c^2 }}\left( {\frac{{\partial f\left( {t - r/c} \right)}} {{\partial
t}}} \right)^2.
\end{equation}
Using this, and the expression for the potential equation (\ref{42}), the radiated power, (\ref{44}), is
\begin{equation} \label{46}
P = \frac{G} {{4\pi c^5 }}\mathop{{\int\!\!\!\!\!\int}\mkern-21mu \bigcirc}
 {d\Omega_{\mathbf{r}} } \left( {{\mathbf{\hat r}}^T {\mathbf{\dddot I}}\,{\mathbf{\hat r}}} \right)^2
\end{equation}
where $\mathbf{I}$ is the matrix of retarded quadrupole moments:
\begin{equation} \label{47}
I_{i,j}  = \int {d^3 {\mathbf{r'}}} r'_i r'_j \rho_m \left( {{\mathbf{r'}},t - r/c} \right)\,.
\end{equation}
The orientation average in equation (\ref{46}) can be computed as follows. Since $ \mathbf{I} $ and $ \mathbf{\dddot I} $ are
symmetric they can be diagonalised by a similarity transform, ${\mathbf{\dddot I}} = {\mathbf{U}}^T {\mathbf{\Lambda U}} $, where
$\mathbf{\Lambda}$ is diagonal and $\mathbf{U}$ is a real orthogonal matrix representing a proper rotation. It follows that the
integration in equation (\ref{46}) can be performed in the rotated coordinate system, ${\mathbf{r}} \to {\mathbf{Ur}}$, with $
\mathbf{\dddot I} $ replaced by $\mathbf{\Lambda}$:
\begin{equation} \label{48}
P = \frac{G} {{c^5 }}\left\langle {\left( {{\mathbf{\hat r}}^T {\mathbf{\Lambda \hat r}}} \right)^2 } \right\rangle\,,
\end{equation}
where $\left\langle {} \right\rangle $ denotes the orientation average. (The Jacobian for the transformation is $ \det \left(
{{\mathbf{U}}^{ - 1} } \right) = \det \left( {\mathbf{U}} \right) = 1 \,.$) Writing $ {\mathbf{R}} \equiv {\mathbf{\hat r\hat r}}^T =
{\mathbf{R}}^T \,,$ one has
\begin{equation} \label{49}
\begin{split}
\left( \mathbf{\hat r}^T \mathbf{\Lambda} \mathbf{\hat r} \right)^2
    &= \trace \left( \left( \mathbf{R\Lambda} \right)^2 \right) \\
    &= \sum\limits_{ijkl} R_{ij} \Lambda_{jk} R_{kl} \Lambda_{li}
    = \sum\limits_{ij} {R_{ij}}^2 \Lambda_{jj} \Lambda_{ii}\,.
\end{split}
\end{equation}
In a Cartesian basis $ \langle {R_{ij}}^2 \rangle$ has only two distinct values: one for all the diagonal terms, and one for all the
off-diagonal terms. Therefore
\begin{equation} \label{50}
\begin{split}
\left\langle {\left( {{\mathbf{\hat r}}^T \mathbf{\Lambda} {\mathbf{\hat r}}} \right)^2 } \right\rangle
    &=  \left\langle {{R_{aa} }^2 } \right\rangle \sum\limits_i {{\Lambda_{ii}}^2 }
        + \left\langle {{R_{ab} } ^2 } \right\rangle \sum_{\substack{  i,j \\ i \ne j}} \Lambda_{ii} \Lambda_{jj} \\
    & =  \left\langle { {R_{aa} } ^2  - {R_{ab} } ^2 } \right\rangle \trace
        \left( {{\mathbf{\Lambda }}^2 } \right)
        + \left\langle { {R_{ab} } ^2 } \right\rangle \trace ^2 \left( {\mathbf{\Lambda }} \right)
\end{split}
\end{equation}
where $ a \ne b\,.$ One finds by explicit calculation that $ \left\langle { {R_{aa} }^2 } \right\rangle  = 1/5$ and $\left\langle {
{R_{ab} } ^2}\right\rangle = 1/15\,.$ Using this and that the trace is insensitive to rotations, one finally obtains
\begin{equation} \label{51}
P = \frac{G} {{15c^5 }}\left( {2\trace \left( {{\mathbf{\dddot I}}^2 } \right) + \trace ^2 \left( {{\mathbf{\dddot I}}} \right)}
\right)\,.
\end{equation}
By contrast, the GR result is \citep{L&L,MTW}:
\begin{equation} \label{52}
P_{GR}  = \frac{G} {{5c^5 }}\trace \left( \dddot{ \reducedI} ^2 \right)
\end{equation}
where $\reducedI$ is the reduced quadrupole moment tensor $ \equiv \mathbf{I} - \mathbf{1} \trace \left( \mathbf{I} \right) /3\,.$
Using that
\begin{equation} \label{53}
\begin{split}
\trace \left( {\dddot{\reducedI}^2 } \right)
    &= \trace \left( {\left( {{\mathbf{\dddot I}} -
        \frac{{\mathbf{1}}} {3} \trace \left( {{\mathbf{\dddot I}}} \right)} \right)^2 } \right) \\
    &= \trace \left( {{\mathbf{\dddot I}}^2 -
        \frac{2}{3} \mathbf{\dddot I} \trace \left( {{\mathbf{\dddot I}}} \right) + \frac{1}{9} \mathbf{1} \trace ^2
        \left( {{\mathbf{\dddot I}}} \right)} \right) \\
    &= \trace \left( {{\mathbf{\dddot I}}^2 } \right) - \frac{1} {3}\trace ^2 \left( {{\mathbf{\dddot I}}} \right)
\end{split}\,,
\end{equation} the GR and PV results can be combined into a single expression
\begin{equation} \label{54}
P\left( \lambda  \right) = \frac{G} {{60c^5 }}\left( {3\left( {3 + \lambda } \right)\trace \left( {{\mathbf{\dddot I}}^2 } \right) +
2\left( {1 - 3\lambda } \right)\trace ^2 \left( {{\mathbf{\dddot I}}} \right)} \right)
\end{equation}
where, as before, the values $ \lambda  = 1,-1/3 $ correspond to GR and PV respectively.
%
\subsection{Orbital decay of a binary system}
%
The components of the quadrupole moment tensor for a pair of objects in mutual elliptical orbit in the $x,y$ plane are \citep{Peters}
\begin{equation} \label{55}
\dddot {\mathbf{I}} =  - \beta \left( 1 + e\cos \psi \right)^2 \mathbf{M}
\end{equation}
where $\mathbf{M}$ is a 3x3 matrix with components
\begin{equation} \label{155}
\begin{split}
M_{xx} &= - 3e\sin \psi \cos ^2 \psi  - 2\sin 2\psi \\
M_{yy} &= e\sin \psi \left( {1 + 3\cos ^2 \psi } \right) + 2\sin 2\psi \\
M_{xy} &= M_{yx} =  2\cos 2\psi  + e\cos \psi \left( {3\cos ^2 \psi  - 1} \right) \\
M_{iz} &= M_{zi} = 0
\end{split}
\end{equation}
where $ i \in \{x,y,z\}\,,$ and
\begin{equation} \label{56}
\beta ^2  =  \frac{{4G^3 m_1^2 m_2^2 \left( {m_1  + m_2 } \right)}} {{a^5 \left( {1 - e^2 } \right)^5 }}\,,
\end{equation}
and where the $m_i$ are the respective masses, $\psi$ is the phase, $a$ is the semi-major axis, and $e$ is the eccentricity of the
orbits. Use of equations (\ref{55}) and (\ref{155}) gives
\begin{equation} \label{57}
\trace \left( {{\mathbf{\dddot I}}^2 } \right) = \beta ^2 \left( {1 + e\cos \psi } \right)^4 \left( {e^2 \sin ^2 \psi  + 8\left( {1 +
e\cos \psi } \right)^2 } \right)
\end{equation}
and
\begin{equation} \label{58}
\trace ^2 \left( {{\mathbf{\dddot I}}} \right) = \beta ^2 \left( {1 + e\cos \psi } \right)^4 e^2 \sin ^2 \psi\,.
\end{equation}
Substitution of these into equation (\ref{54}) gives that the total power is
\begin{equation} \label{59}
P\left( \lambda,\psi  \right) = \frac{{G\beta ^2 }} {{60c^5 }}\left( {1 + e\cos \psi } \right)^4 X\left( \lambda,\psi  \right)
\end{equation}
where
\begin{equation} \label{159}
X = \left( {11 - 3\lambda } \right) e^2 \sin ^2 \psi
      + 24\left( {3 + \lambda } \right)\left( {1 + e\cos \psi } \right)^2\,.
\end{equation}
Since interest is restricted to secular effects, it is safe to work with the power averaged over an orbital cycle, $\tau$, say: $
\overline {P}\left( \lambda  \right)  =  \int\limits_0^\tau {dt\,} P\left( \lambda,\psi  \right)/\tau $. Using that, for any function
$f$,
\begin{equation} \label{60}
\int\limits_0^\tau  {dt} f\left( \psi  \right) = \int\limits_0^{2\pi } {d\psi } {{f\left( \psi  \right)} \mathord{\left/
 {\vphantom {{f\left( \psi  \right)} {\dot \psi }}} \right.
 \kern-\nulldelimiterspace} {\dot \psi }}
\end{equation}
and that the rate of change of phase of an elliptical orbit is
\begin{equation} \label{61}
\dot \psi  =  k\left( {1 + e\cos \psi } \right)^2
\end{equation}
where $k$ is a constant, the time-average of equation (\ref{59}) is
\begin{equation} \label{62}
\overline {P}\left( \lambda  \right)  = \frac{{G\beta ^2 }} {{60c^5 }}
    \frac{\int\limits_0^{2\pi } {d\psi } \left( {1 + e\cos \psi } \right)^2 X\left( \lambda,\psi  \right)}
    {\int\limits_0^{2\pi } {d\psi } \left( {1 + e\cos \psi } \right)^{ - 2} }\,.
\end{equation}
Performing the integrations and recalling equation (\ref{56}) one obtains
\begin{equation} \label{63}
\overline {P}\left( \lambda  \right)  =  \frac{{32}} {5}\frac{{G^4 m_1^2 m_2^2 \left( {m_1  + m_2 } \right)}} {{a^5 c^5 }}f\left(
{\lambda ,e} \right)
\end{equation}
where
\begin{equation} \label{64}
f\left( {\lambda ,e} \right) =  \frac{{\left( {227 + 69\lambda } \right)e^4  + 4\left( {443 + 141\lambda } \right)e^2  + 192\left( {3
+ \lambda } \right)}} {{768\left( {1 - e^2 } \right)^{7/2} }}
\end{equation}
is the model-dependent `enhancement factor', with coefficient chosen so that $ f\left( {1,0} \right) =  1 $. It is easily verified
that the familiar (GR) result of Peters and Mathews is recovered when $\lambda = 1$. The ratio of the PV to GR radiated powers,
\begin{equation} \label{65}
\frac{\overline {P} \left( - 1/3 \right)} {\overline {P}\left( 1 \right)}
    =  \frac{f\left(  - 1/3 , e \right)}{f\left(  1 , e \right)}
    = \frac{1} {2}\left( \frac{51e^4  + 396e^2  + 128} {37e^4  + 292e^2  + 96} \right)
\end{equation}
is plotted in Fig. 6 as a function of the eccentricity. The value of the ratio in the limit of perfectly circular orbits ($e = 0$) is
exactly 2/3, and changes only by 1.5\% across the full range of eccentricity. That the ratio is less than one is to be expected given
the reduced number of polarizations available to the vacuum $K$-field compared with the two independent polarizations of weak-field GR
radiation. This is in accord with divergence between GR and other theories that have different rules governing generation of radiation
\citep{Will}.
%
\subsection{Test against observations of orbital decay of a binary system}
%
Using the Kepler results that the energy of the binary system is $ E = - Gm_1 m_2  /2a $, and its period is
    $ \tau_b = {{2\pi a^{3/2}}
    \mathord{\left/ {\vphantom {{2\pi a^{3/2} } {\sqrt {G\left( {m_1  + m_2 } \right)} }}} \right.
    \kern-\nulldelimiterspace} {\sqrt {G\left( {m_1  + m_2 } \right)} }}
    $,
one has $ \dot \tau_b /\tau_b =  - 3P/2E$ and therefore \citep{Damour,Wagoner}
\begin{equation} \label{66}
\dot \tau_b  =    - \frac{{192\pi }} {5}\nu \left( \frac{2\pi G\left( {m_1  + m_2 } \right)} {\tau_b c^3 } \right)^{5/3}
    f\left({\lambda ,e} \right)
\end{equation}
where
$ \nu  = {{m_1 m_2 } \mathord{\left/
 {\vphantom {{m_1 m_2 } {\left( {m_1  + m_2 } \right)^2 }}} \right.
 \kern-\nulldelimiterspace} {\left( {m_1  + m_2 } \right)^2 }}\,,$
and where the period has been used to eliminate the semi-major axis. Estimates for the masses may be obtained from the rate of
periastron advance, $ \dot \omega $, and the time-dilation parameter $\gamma$ in the case there is apsidal motion \citep{Blandford}.
It is important that though inference of these quantities from observation is potentially model dependent, they are computed using the
static metric of the center of mass system, in which case the PPN expansion of PV is known to give the same result as GR to the
required order \citep{Puthoff}. Consequently, the PV-dependent observational inference of these quantities, and therefore $m_1$ and
$m_2$, is the same as that of GR. Then, if one regards (66) as a theoretical prediction for the decay rate $ \dot \tau_b $ in terms of
the observationally determined quantities $ \tau_b $, $m_1$ and $m_2$, it is concluded that the prediction from PV differs from that
by GR by the factor given in equation (\ref{65}) of approximately 2/3. But since the observed decay rate of PSR 1913 + 16 agrees with
the prediction of GR to better than 1\%, with more recent research concerned with checking relativistic corrections and other
refinements of the analysis presented here \citep{Taylor}, it follows that PV is incompatible with observation.
%
\begin{figure}
\label{fig6}
\plotone{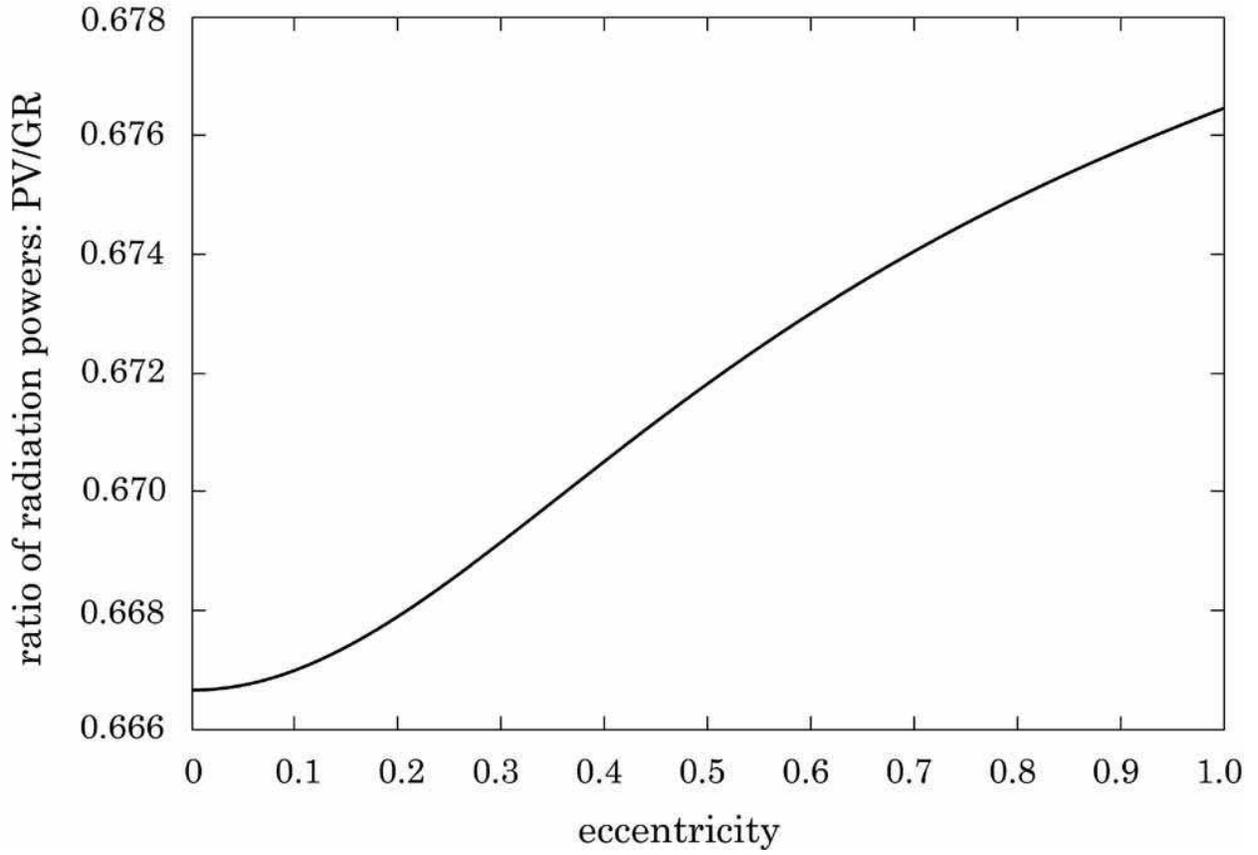} 
    \caption{Ratio, PV/GR, of gravitational radiation enhancement factors. The original definition by Peters and Mathews is the ratio of
radiation power from a binary system with elliptical orbits to that from a system with circular orbits.}
\end{figure}
%
%
\subsection{Summary of gravitational radiation findings}
%
PV predicts gravitational radiation with the following properties
\begin{enumerate}
\item Radiation is polarized along the direction of propagation; there is just one polarization degree of freedom.

\item The orbital decay due to radiation of a binary system is always close to 2/3 that predicted by GR, with the exact relation as a
function of eccentricity given by equation (\ref{65}), and is plotted in Fig. 6.

\item This prediction is at variance with the observational data for PSR 1913 + 16.
\end{enumerate}
%
\section{Developments}
\label{sec:developments}
\subsection{Anisotropic polarizability and the Yilmaz theory}
\label{sec:anisotropy}
%
At the time of Dicke's writing, his theory passed the then four standard tests of GR. Dicke did not investigate the cosmological
predictions of the theory, which are reported here for the first time. The major shortcoming of PV reported immediately above - its
failure to predict the correct (observed) orbital decay rate of a binary system due to radiation - was unknown to Dicke. The factor of
2/3 disagreement between PV and GR possibly can be attributed to the reduced number of polarization states permitted to the radiation.
Depending on the details, a theory accommodating vacuum anisotropy has the potential to accommodate two independent transverse
polarization degrees of freedom of radiation, thereby bringing about agreement with the binary pulsar data (and GR). In fact Dicke
briefly considered the possibility of an anisotropic version of the theory as plausible, but argued in favor of the scalar version,
since it was sufficient to explain the standard test data available at that time \citep{Dicke1}.

The Yilmaz theory is a candidate for the generalized - anisotropic - version of PV. As remarked in the introduction, the Yilmaz theory
shares with PV the electromagnetic motivation, and, in the case that only $T_{00}$ is non-zero, the same form for the metric. This
suggests that PV may turn out to be a truncated version of the Yilmaz theory, in the same sense that a theory of electromagnetic
interaction would be truncated by keeping $\phi$ but ignoring $\mathbf{A}$. Since the Yilmaz theory has the full compliment of 10
functional degrees of freedom in the metric, it has the potential to resolve the problem of the reduced radiation rate suffered by PV.
To date there have been no published studies of the cosmological predictions of the Yilmaz theory, so it would be interesting to
pursue this matter further in a future report. Since the Friedmann equation is predicated on isotropy and homogeneity on the
cosmological scale, it is conceivable that such a development will affect only the compatibility with the binary pulsar data through
the polarization states available to radiation, without affecting the cosmological predictions of PV reported above. However, no
conclusions are possible based on these very broad arguments.
%
\subsection{Vacuum contribution}
%
It is recalled that for the Einstein equation with a Cosmological term to retain its tensor form, it is necessary that the vacuum
action must have the form $ {\text{constant}} \times \sqrt { - g} $. With this, the FLRW flat-space metric (i.e. with line element
equation (\ref{12})) gives that the vacuum contribution appears in the Lagrangian density as $ {\text{constant}} \times a\left( t
\right)^2 $ (equation (\ref{5})). The PV field $K$, however, is a scalar, and therefore its (Euler-Lagrange) equation of motion is a
scalar. It follows that there is no such constraint on the vacuum contribution to the PV action. In principle therefore, other vacua
may be contemplated, with potentially radical consequences for the resulting cosmology. In the absence of a solid mathematical basis
for a particular form for the vacuum term, some other argument in favor of a particular alternative would have to be provided in order
to proceed further. It is noteworthy though that PV cosmology is in closest agreement with observation when the vacuum term is close
to zero. Given this, and the arbitrariness of the function dependence of the vacuum term on the scale factor, it could be argued that
the vacuum term be expunged from PV altogether.

Alternatively, one might argue that with no constraints on the vacuum dependence on the scale factor there is some justification for
regarding the total energy term ($ \Omega_\Sigma $ in equation (\ref{15})) as a novel, negative-energy vacuum term. That is, equation
(\ref{5}) could be replaced with
\begin{equation} \label{67}
L_{vac}  \to  - \int {d^4 x\,a^2 \rho_v \left( a \right)};\quad \rho_v \left( a \right) = - \rho_v /a^2
\end{equation}
where now $ \rho_v $ is a positive constant, which would then generate the Friedmann equation
\begin{equation} \label{68}
\frac{\lambda } {H_0^2 }\left( {\frac{da} {d\tau }} \right)^2  = \frac{\Omega_m } {a} + \frac{\Omega_r } {a^2 } - \Omega _v\,.
\end{equation}
The advantage of this view is that there is now no arbitrary overall energy - the total energy is now zero. Of course, the
arbitrariness in $ \Omega_\Sigma $ has just been subsumed into the arbitrariness of the vacuum.
%
\subsection{PV with a spatially-varying $K$-field}
%
Recalling that the PV action is as if the matter and fields are in a curved space-time with line element (\ref{PVLineElement}), (of
which the flat-space FLRW line-element of equation (\ref{7}) is obviously, therefore, a special case), it is observed that Dicke's
scalar PV is committed to an isotropic space. In the case that $K$ is just a function of time - as it is here - then equation
(\ref{7}) is the FLRW flat space metric (with proper time $\tau$ then given by $ d\tau  = {{dt} \mathord{\left/
 {\vphantom {{dt} {\sqrt {K\left( t \right)} }}} \right.
 \kern-\nulldelimiterspace} {\sqrt {K\left( t \right)} }}{{ = dt} \mathord{\left/
 {\vphantom {{ = dt} {a\left( t \right)}}} \right.
 \kern-\nulldelimiterspace} {a\left( t \right)}}
$ ). It is obvious from the form equation (\ref{7}) that the other FLRW space-geometries (spherical and hyperbolic) are not
accommodated PV. The line elements corresponding to those geometries were first derived independent of GR (see for example
\citet{Tolman}) based upon Milne's Cosmological Principle, which, if regarded as an axiom, dictates that PV cosmology entertain only
the flat-space metric.

It is noteworthy that the specific form of the three FLRW line elements (plus, of course, the Friedmann equation) can be generated
from `within' GR, i.e. without direct appeal to the Cosmological Principle, and without enforcing any symmetry by hand. For example,
it sufficient (in order to generate all FLRW three metrics), that that the matter is required to be static and uniformly distributed
(in whatever space it exists), and that the metric can be written in the form
\begin{equation} \label{69}
ds^2  = c^2 dt^2  - f\left( {t,{\mathbf{x}}}
\right)d{\mathbf{x}}^2
\end{equation}
Given that this derivation of the FLRW metrics is, in some sense, achieved more `directly' by the theory of gravity (rather than by a
symmetry principle), it could be argued that this is the physically superior of the two routes to the metric. From that standpoint, a
similar procedure should be used to decide on the cosmological metrics appropriate to PV. To that end, alternatives to the flat space
assumption might arise from an `effective' cosmologically-appropriate version of the PV action, equation (\ref{2}), that does not, a
priori, exclude spatial variation of $K$:
\begin{equation} \label{70}
\begin{split}
L =  - \int {d^4 x}
    \Biggl\{
      \frac{c^4 } {32\pi G} & \left[ \left( {\frac{\nabla K} {K}} \right)^2
        - \left( \frac{1}{c} \frac{\partial K} {\partial t} \right)^2  \right]
     { + \frac{\rho_m}{\sqrt K } + \frac{\rho_r}{K} + K\rho_v}
    \Biggr\}\,.
\end{split}
\end{equation}
Variation of $K$ gives the Euler equation
\begin{equation} \label{71}
\frac{\nabla ^2 K}{K} - \left( \frac{\nabla K}{K} \right)^2
    - \frac{K}{c^2} \frac{\partial ^2 K}{\partial t^2}
    = \frac{16\pi G}{c^4}\left( \rho_v
     - \frac{\rho_m } {2\sqrt K } - \frac{\rho_r } {K} \right)
\end{equation}
where the $\rho_i$ are constant. In principle, given plausible boundary conditions for, say, $K$ and $ \partial K/\partial t $
everywhere at $t = 0$, solutions to equation (\ref{71}) should provide cosmologically acceptable functions of $K$ that vary in space
and time and which will generate PV analogues of the FLRW metrics.
%
\subsection{Relation between spatially-varying $K$-field and vacuum and overall energy densities}
%
In the Hamiltonian corresponding to the action in equation (\ref{70}) the spatial gradient term contributes positively, with the same
sign as the matter and radiation terms (the sign of the vacuum term is not fixed). Consequently any spatial variation in the metric
increases the overall system energy. This is to be contrasted with the effect of the spherical and hyperbolic FLRW metrics in GR,
which contribute terms of opposite signs to the Friedmann equation for the scale factor $a\left(t\right)$. Since agreement with
observation in a flat-space zero vacuum cosmology already requires that PV have non-zero positive energy (i.e. $ \Omega_\Sigma \approx
0.3 $ identified above), it follows that the introduction of space-dependent forms for $K$ cannot be used to offset this requirement.
In other words, PV without a contribution from the vacuum cannot have zero overall energy in any space-time.
%
\subsection{Possibility of a radiation era}
%
Analysis above showed that flat-space PV has no baryon, lepton, or radiation eras. Though the baryon and lepton eras were shown to be
orders of magnitude beyond the reach of PV, the same was not true of the radiation era, which was discounted on the basis of an
inferred age of 6.6 Gyr or less. It is conceivable that a spatially varying $K$-field satisfying equation (\ref{71}) with
appropriately-chosen boundary conditions could satisfy the observational constraints on age etc, and bring a radiation era within the
scope of PV by a small adjustment of the metric at very early times (yet without significantly changing the later evolution).
%
\subsection{Gravitational redshift and Yilmaz stars}
%
With reference to equation (\ref{71}), the vacuum behaviour of the $K$-field near a massive object (ignoring Cosmological influences)
is given by
\begin{equation} \label{72}
{{\nabla ^2 K} \mathord{\left/
 {\vphantom {{\nabla ^2 K} K}} \right.
 \kern-\nulldelimiterspace} K} - \left( {{{\nabla K} \mathord{\left/
 {\vphantom {{\nabla K} K}} \right.
 \kern-\nulldelimiterspace} K}} \right)^2  = 0\,.
\end{equation}
Setting $ \psi  = \log K $
 gives the Laplace equation, $
\nabla ^2 \psi  = 0 $, from which it is inferred that
\begin{equation} \label{73}
\psi  = \sum\limits_{l = 0}^\infty  {\sum\limits_{m =  - l}^l
{\left( {a_{lm} r^l  + b_{lm} r^{ - l} } \right)Y_{lm} \left(
{\theta ,\phi } \right)} }
\end{equation}
where the $ Y_{lm} \left( {\theta ,\phi } \right) $ are spherical harmonics, and where $a_{lm} $ and $b_{lm} $ are arbitrary constants
to be determined by the boundary conditions. But far from the source of the field $K$ must return to a constant value, so all the $
a_{lm} $ and $ b_{0m} $ must be zero, and therefore
\begin{equation}
\label{74}
K = \prod\limits_{l = 1}^\infty  {\prod\limits_{m =  - l}^l {\exp
\left( {{{b_{lm} Y_{lm} \left( {\theta ,\phi } \right)}
\mathord{\left/
 {\vphantom {{b_{lm} Y_{lm} \left( {\theta ,\phi } \right)} {r^l }}} \right.
 \kern-\nulldelimiterspace} {r^l }}} \right)} }
\end{equation}
where the remaining $ b_{lm} $ can be determined by the conditions at the surface of the source. The important feature of equation
(\ref{74}) is that, regardless of the mass distribution, $K$ is everywhere finite outside the mass. Recalling the PV line element,
(\ref{PVLineElement}), it is seen that the coefficient of the $dt^2$ term in the line element is never zero, from which it is deduced
that PV does not admit event horizons outside the matter distribution; PV predicts that light from the surface of any star, however
massive, cannot be infinitely redshifted, i.e., a star can grow to an arbitrary density without disappearing from view. Hence it is
concluded that PV does not admit black holes. This statement does not address the issue of stability; properly applied with an
appropriate equation of state, PV may or may not concur with GR predictions for instability and collapse. At this point the only
certain (but important) difference is that any collapsed star will nonetheless remain visible, albeit with an arbitrarily large
redshift. As a result of this difference with GR, a more thorough appraisal of the validity of PV might need to take into account the
possibility that GR-designated cosmological redshifts may have a significant gravitational component. Since, as pointed out in section
\ref{sec:anisotropy}, the exponential form of the vacuum metric, equation (\ref{74}), is common to PV and the Yilmaz theory, the lack
of black holes is a prediction common to both theories.
%
\subsection{Relation to the quasi-steady-state cosmology}
%
Several authors have presented arguments in favor of a quasi-steady-state cosmology (QSSC) that has no big bang
\citep{Burbidge1,Narlikar,Volkov}. A presentation of both sides of the QSSC-GR debate may be found in \citet{Burbidge2} and
\citet{Albrecht}, respectively. Lacking BBN, they have argued that observed abundances of the light elements are the result of stellar
rather than big bang nucleosynthesis - a feature of potential relevance to PV. Also in common with PV, QSSC also lacks an era of
radiation domination. QSSC claims that the CMB is thermalized starlight, thermalisation occurring principally in the most contracted
phase of an oscillating cosmology. This component of the QSSC hypothesis is potentially relevant exclusively to the oscillating PV
cosmologies, including the particular example discussed above and shown in Fig. 4 (which happens to have the same oscillation period,
of around 100 Gyr, as that preferred by QSSC). The degree of thermalisation of light from a previous cycle will depend on the rate and
depth of the contraction, the details of which differ in QSSC and PV. Further investigation would be necessary to test the viability
of this idea within PV.
%
\section*{Acknowledgements}
%
The author is very grateful to H.E. Puthoff and S. Little for fruitful and enjoyable discussions and suggestions.
%

\end{document}